\documentstyle[prd,preprint,aps]{revtex}

\newcommand{\be}{\begin{equation}}
\newcommand{\ee}{\end{equation}}
\newcommand{\bea}{\begin{eqnarray}}
\newcommand{\eea}{\end{eqnarray}}

\newcommand{\V} {{\cal V}_n}
\newcommand{\Y} {{\overline \psi}}

\begin{document}

\reversemarginpar
\tighten

\title{Nearly Degenerate dS Horizons
from a 2-D Perspective} 

\author {A.J.M. Medved}

\address{
Department of Physics and Theoretical Physics Institute\\
University of Alberta\\
Edmonton, Canada T6G-2J1\\
E-Mail: amedved@phys.ualberta.ca\\}

\maketitle

\begin{abstract}

Some recent studies have implied that quantum fluctuations 
will prevent a near-extremal black hole from ever
attaining a state of precise extremality.  
In this paper, we consider the analogous situation
for the scenario of a nearly degenerate Schwarzschild-de Sitter
black hole (of arbitrary dimensionality).
For this purpose, we utilize a holographic type of 
duality between the solutions of interest and 
the near-massless sector of (two-dimensional) Jackiw-Teitelboim theory.
After explicitly demonstrating this duality, we go
on to argue that, on the basis of one-loop
considerations, a similar censorship applies
in  this de Sitter context. That is, quantum
back-reaction effects will conspire to 
prohibit a non-degenerate Schwarzschild-de Sitter spacetime
 from continuously evolving into a
degenerate ({\it i.e.}, Nariai) black hole solution.

\end{abstract}

\section{Introduction}

From a theoretical perspective,
black holes
provide us with an intriguing   ``laboratory''
for examining many aspects of quantum and semi-classical gravity.
With regard to such investigations,  the thermodynamic properties 
associated with  black
hole horizons   have played an especially 
prominent role.   
In particular, black holes are commonly attributed
a temperature  that is proportional to the horizon surface gravity
and an entropy that is proportional to the horizon surface area 
\cite{wald}.
The initial motivation for these identifications
was an observed analogy between black hole horizon mechanics
and the thermodynamic first and second laws \cite{bek,haw}. Nonetheless,
this thermodynamic correspondence is generally given
a physical status on the basis of Hawking radiation \cite{haw2}.
That is, it has been shown that, when quantum effects are considered, 
 black holes emit thermal radiation and, remarkably,
at  the exact value of temperature that
the analogy suggests.
\par
On the other hand,
in a statistical mechanical sense, there are still 
conceptual difficulties with this thermodynamic 
picture. For instance, ``what is the microscopic
origin of the black hole entropy?'' has become
a popular slogan (paraphrasingly speaking) among 
gravitational researchers \cite{CAR}.
However, putting this controversial issue  aside,
one can  see that the entropy (and not only
the temperature) does, indeed, have a clear physical interpretation. 
To elaborate, the black hole horizon is essentially a membrane
that obstructs the flow of information. That is,
an external observer will be unable to access
any degrees of freedom from behind the horizon, and  just such
an obstruction will naturally induce, via coarse-graining effects,
  an entropy of entanglement \cite{BEK2}.
 Indeed,  direct calculations
of this entanglement entropy ({\it e.g.}, \cite{SRE})  
have been known to reproduce the Bekenstein-Hawking
entropic  area law \cite{bek,haw}. 
(Albeit, up to some ambiguity in the proportionality
factor. However, it is expected that the correct factor   
will be reproduced once renormalization effects have
properly been accounted for
\cite{SUUG}.)
\par
Moreover,
there is substantial evidence that  any horizon ({\it i.e.},
any null hypersurface that prevents the accessibility of information)
can be attributed a temperature and entropy in the
same manner as for a black hole ({\it e.g.},\cite{PAD}). 
Although intuitively sensible, this may 
not be {\it a priori} clear for horizons that are observer
dependent; most notably, Rindler horizons (which are  manifestations
of an accelerating observer \cite{BD}) and  de Sitter cosmological
horizons (which are defined in terms of a given observer's
causal patch of de Sitter space \cite{SSV}). On the other hand, 
the  principle of {\it black hole complementarity} \cite{SUSP} implies
that the existence (or lack thereof) of a black hole horizon is 
also an objective  feature of the  spacetime; for instance,
a free-falling observer will detect no horizon at all.
Hence, observer dependence should not, after further consideration,
be that significant of an issue.
\par
Furthermore, our understanding of horizon thermodynamics is expected to
persist in  spacetimes with multiple horizons ({\it e.g.}, charged
and/or spinning black holes and Schwarzschild-de Sitter spacetimes),
provided that each horizon is treated locally \cite{PAD} or as a distinct
physical system 
\cite{TEI}.\footnote{It has   recently been conjectured, however,
that the  temperature in a Schwarzschild-de Sitter spacetime
can be given a global meaning \cite{SHANK}.}   
There is, however, a very important caveat to this statement 
when two (or more) otherwise
distinct horizons become degenerate. For such an occurrence,
 the thermodynamic properties  typically become 
ambiguous and sometimes even  ill-defined (for further discussion and
references, see \cite{me1}).
\par
This formal breakdown in degenerate-horizon thermodynamics
brings us to the problem of
how to interpret the degenerate limit of relevant spacetimes;
most notably, the extremal limit of a charged and/or spinning
black hole and the Nariai limit \cite{NAR} of a Schwarzschild-de Sitter
spacetime. (Since a degenerate  limit  naively  coincides
with a limit of vanishing temperature, this
problem can alternatively be viewed as an ambiguity
in formulating the third law of horizon thermodynamics \cite{WALD2}.)
That such a problem exists is re-enforced when
one considers these degenerate limits from a topological perspective.
That is, the  degenerate solutions (as indicated above) have 
non-trivial qualitative differences in  topology from 
their non-degenerate counterparts ({\it e.g.},\cite{hhr,BOU}). 
\par
The resolution of this degenerate-limit problem is, in some
(perhaps naive) sense, really
quite simple. One needs only to conjecture that
a non-degenerate solution is prohibited from
continuously evolving into a degenerate one
(and {\it vice versa}) \cite{hhr}. That this should intuitively be the case
follows from the topological distinctions mentioned above.
(This still does not resolve the related problem of
how  should the thermodynamic properties in a degenerate spacetime
be defined. We will not, however, be addressing this
issue in the current paper.\footnote{But for further discussion and references
in an extremal black hole context, see \cite{me1}.}  Rather, our focus
is on the  ``near-degenerate'' sector of a  particular 
theory.) To physically motivate this  conjecture, one 
still requires a mechanism for enforcing the proposed censorship. 
Fortunately, it appears that quantum  effects   efficiently
and naturally serve this purpose. Indeed, recent
studies in the context of black
hole spectroscopy   have directly implied
that quantum fluctuations will prevent
a charged and/or spinning black hole from
ever reaching a precise state of extremality \cite{das,US}.
Also recently, a similar outcome has  been obtained by the current
author via independent
means \cite{new,MED}. As this latter approach is central to
the present analysis, let us give a point-by-point recount
of the relevant steps and results. \\ 
{\it (i)}  We started off with  a  Reissner-Nordstrom 
({\it i.e.}, charged and static)
black hole  in an asymptotically flat spacetime.
(The number of dimensions was initially four \cite{new}
but later arbitrary \cite{MED}.)
The focus was on the near-extremal sector;
in particular, the {\it deviations} (from extremality) in
the black hole  entropy and 
temperature 
were determined as  functions of the {\it deviation} in  
the mass (with the charge regarded as a fixed parameter). \\
{\it (ii)} Subjecting the ``fundamental'' action  to a
 procedure of reduction and field
reparametrization, we  obtained a two-dimensional
effective model.
 (The motivation being that
quantum back-reaction effects, which are ultimately of interest,
 can be much more easily 
addressed in the lower-dimensional theory.) 
Then following \cite{x5,fnn}, we expanded  the two-dimensional action 
about the extremal configuration. The resulting theory,
which should effectively describe the Reissner-Nordstrom
near-extremal sector, turned out to be a two-dimensional
anti-de Sitter action; that is, the Jackiw-Teitelboim
model \cite{jt}. \\
{\it (iii)} A  duality was re-established \cite{x5,fnn}
between the near-extremal Reissner-Nordstrom black
hole and the near-massless sector of
Jackiw-Teitelboim theory. That is, after appropriate
identifications, the thermodynamic
deviations (as discussed above) were shown to coincide
exactly with the thermodynamics of the two-dimensional  black
hole solutions.  \\
{\it (iv)}  The next phase of the analysis
concentrated on
 first-order quantum effects.
Here, we began by incorporating
the simplest possible matter field having a higher-dimensional
pedigree; namely, a massless scalar field that is
minimally coupled to the fundamental gravity
theory. The  same procedure of dimensional reduction,
field reparametrization and expansion  was then repeated
for the revised theory. Interestingly, we found that the 
revised effective
action  mimics the theory of a
dimensionally  reduced  BTZ black hole\footnote{The
BTZ black hole refers to special solutions of 2+1-dimensional
anti-de Sitter gravity that exhibit all the properties
of black holes \cite{btz}. The dimensionally reduced
BTZ model was first discussed in \cite{ao}.}
with minimally coupled matter. \\
{\it (v)} Eventually,  attentions were focused on the  
quantum-corrected form of the surface gravity  or (equivalently) 
the  temperature. 
The desired result, at the one-loop level,
could be directly extracted from a prior study  on
the thermodynamics of  
dimensionally reduced BTZ black holes  
\cite{me1} (also, \cite{REV}).\footnote{Actually, \cite{me1} used a
form of  dimensionally reduced (one-loop)  action  that
 has recently  been  criticized on technical
grounds \cite{FRO,ZEL}. If the action in question does require
(yet unknown) corrections, then it is certainly  clear that
 any quantitative results would have to
be closely scrutinized. However,  we suggest  that the coarse,  qualitative
features of the  spacetime would be less sensitive to
any such corrections.} 
Working under the premise that a consistent 
theory must have a non-negative surface gravity \cite{aht},
we were able to establish a {\it finite} lower bound on the black hole mass.
 That is,  quantum back-reaction
effects will prevent the (two-dimensional) black hole from completely
evaporating; rather, the system will ``freeze'' at a
non-vanishing value of mass. \\
{\it (vi)} In view of the prior outcome
{\it and} the previously established duality, we ultimately argued
that quantum back-reaction effects will 
prevent  the Reissner-Nordstrom black hole from ever reaching a state
of precise extremality. That is, a non-extremal Reissner-Nordstrom black hole
will be unable to continuously evolve into an extremal solution
(and {\it vice versa}). 
\par
The objective of the current paper is to extend
the above treatment to the scenario of a nearly degenerate
Schwarzschild-de Sitter black hole spacetime. In this
case, the two horizons represent the (inner) black hole
horizon and the (outer)  cosmological horizon of an
asymptotically de Sitter spacetime.  Also of interest, the point
of degeneracy corresponds to the so-called
Nariai black hole solution  \cite{NAR}, which can equivalently
be regarded as the Schwarzschild-de Sitter solution
of maximal mass. For motivation, let us point out that asymptotically
de Sitter spacetimes have experienced a recent surge
in interest; thanks to, for instance, astronomical observations \cite{ASTRO},
possible holographic dualities \cite{STR1} and quasi-de Sitter inflationary
scenarios \cite{DAN}. (For a general overview, see \cite{BOU}.)
\par
The rest of the paper is organized as follows. Section
2 introduces the  ``fundamental'' theory of interest:
a Schwarzschild-de Sitter spacetime of arbitrary 
dimensionality. Here,  the focus is 
on the thermodynamic properties of near-degenerate
solutions.
In Section 3, we employ a suitable
reduction {\it ansatz}, as well as other manipulations, to
  derive an effective two-dimensional
model of the near-degenerate sector.
We go on to demonstrate a clear thermodynamic
duality between this effective description
and the originating theory. (Some of 
this section  follows parts of \cite{x5},
although the perspective of the cited
paper  differs substantially from the present one.) 
 In Section 4, we consider one-loop
corrections and present an argument as
to why back-reaction effects would
prevent a  degenerate
state from ever being reached.
Finally, Section 5 contains a brief summary and
discussion.

\section{The Fundamental  Theory}

\par
Let us formally  begin  by  writing down
the de Sitter gravitational action for a spacetime of
arbitrary dimensionality ($d=n+2 >3$): 
\be
I^{(n+2)}={1\over 16\pi l^n}\int d^{n+2}x\sqrt{-g^{(n+2)}}\left[
R^{(n+2)}-2\Lambda\right],
\label{1}
\ee
where $l^n$ is the $n$+2-dimensional Newton constant (with $l$ being
the analogue to the Planck length) and 
$\Lambda$   is the positive cosmological constant.
Note that
\be
\Lambda={n(n+1)\over 2L^2},
\label{1.5}
\ee
where $L$ is the  curvature radius of de Sitter
space.   
\par
If we assume no electrostatic charges, then the
unique static solution of Eq.(\ref{1}) can be
written as follows:
\be
ds_{n+2}^2= - h(r)dt^2
+{1\over  h(r)}dr^2 +r^2d\Omega_n^2,
\label{2}
\ee
\be
h(r)=1-{16\pi l^n M\over n \V r^{n-1}} -{r^2\over l^2} ,
\label{2.5}
\ee
where $M$  represents the conserved  mass \cite{BDM}  and $d\Omega_n^2$ is
an $n$-dimensional spherical hypersurface with
volume  $\V=2\pi^{{n+1\over 2}}/
\Gamma\left({n+1\over 2}\right)$.
\par
The above solution is that of a (non-degenerate) Schwarzschild-de Sitter
black hole  provided that the mass observable is
constrained according to
\be
0 < M < M_0\equiv{n\over (n+1)}{\V\over 8\pi l^n}
\left[\left({n-1\over n+1}\right)L^2\right]^{n-1\over2}.
\label{2.75}
\ee
For $M$  within these limits, one can readily
verify the existence of exactly two  positive, real
and distinct horizons; which can be  located by solving the polynomial
$h(r=r_H)=0$. The smaller of this pair (to be denoted as
$r_H=r_-$) is a Schwarzschild-like
black hole horizon, while the larger is the de Sitter
cosmological horizon (which will be denoted as $r_H=r_+$). 
\par
Of special interest are the upper and lower limits
in Eq.(\ref{2.75}). Firstly, when $M=0$, there is
no black hole to speak of and the spacetime corresponds to
 pure de Sitter space. In this event, there
is a single (cosmological) horizon at $r_+=L$.
Secondly, when $M=M_0$, the two horizons coincide
and this solution of maximal mass  is often
referred to as the Nariai \cite{NAR} black hole.
More explicitly, by solving for $h(r_H;M=M_0)=0$, one finds
that $r_-=r_+=r_0$, where
\be 
r_0 \equiv \sqrt{{n-1\over n+1}} L.
\label{2.90}
\ee
Let us also point out that either $M<0$ or $M>M_0$ results in a naked
singularity,\footnote{To elaborate, the black
hole horizon disappears when $M<0$ and both horizons vanish
if $M>M_0$.} 
but  these awkward scenarios can (and will) be disregarded
on the grounds of cosmological censorship \cite{WALDT}.   
\par
Since our current focus is on the case of nearly degenerate
horizons, we will now assume that $M$ is close
to, but always less than, the Nariai value of
$M_0$.  To put this notion on a more quantitative
level, let us introduce the following measures
of deviation from degeneracy:
\be
\Delta M \equiv M-M_0, 
\label{2.95A}
\ee
\be
\Delta r_{\pm} \equiv r_{\pm} - r_0,
\label{2.95B}
\ee
while insisting that $|\Delta M|<<M_0$ and $|\Delta r_{\pm}|
<< r_0$. It should be kept in mind that  $\Delta M<0$
must always be imposed. On the other hand, it should
be clear that one of the horizon shifts will  be
positive and the other, negative. By our prior convention,
this means that $\Delta r_+ >0$ and $\Delta r_-<0$.
Moreover,
to lowest order in $\sqrt{\Delta M}$, one can employ
the horizon condition ({\it i.e.}, $h(r=r_H)=0$)
and readily show that
\be
\Delta r_{\pm}=\pm {L\over n+1} \sqrt{{2|\Delta M|\over
M_0}}.
\label{3}
\ee
\par
Let us now consider the  thermodynamic properties
- specifically, the  Bekenstein-Hawking entropy
\cite{bek,haw} and Hawking temperature \cite{haw2,GH} -
of these horizons. (For the generalization of
these properties to a cosmological horizon,
see \cite{GHC,SSV}.)
Here, we will employ the standard prescriptions for
the entropy and temperature (respectively):
\be
S_{H}={A_{H}\over 4\hbar l^n}= {\V r_H^n\over 4\hbar l^n },
\label{4}
\ee
\be
T_{H}={\hbar \over 2\pi} |\kappa_H|= {\hbar\over 4\pi}
\left|{dh\over dr}\right|_{r=r_H}
= \hbar\left|{4(n-1)l^nM\over n\V r_H^n}
-{r_H\over 2\pi L^2}\right|,
\label{5}
\ee
where $A_H$ is the surface area and $\kappa_H$ is
the surface gravity with respect to the horizon at $r=r_H$.\footnote{Note
that, here and throughout, the speed of light and Boltzmann's constant
have been set to unity.}
\par
Some straightforward calculations reveal that,
to lowest order in $\sqrt{|\Delta M|}$ (as appropriate for
the near-degenerate regime of interest),
\bea
\Delta S_{\pm}&\equiv& S_{\pm}(M_0+\Delta M) -S_{H}(M_0)
\nonumber \\
&=&\pm {1\over\hbar} \sqrt{{n \pi \V\over(n-1)}{r_0^{n+1}\over l^n }
|\Delta M|},
\label{7}
\eea
\bea
\Delta T_{\pm}&\equiv& T_{\pm}(M_0+\Delta M)-T_{H}(M_0)
\nonumber \\
&=& 2\hbar{(n-1)\over (n+1)} 
\sqrt{{(n-1)\over n \pi \V}{l^n\over r_0^{n+1}}
|\Delta M|}.
\label{8}
\eea
Note that $S_H(M_0)=\V r_0^n / 4 \hbar l^n $
and $T_H(M_0)=0$.\footnote{Actually, rigorous analysis
indicates that $T=\hbar/2\pi r_0$ for a Nariai black hole
\cite{BOUH}. Nevertheless, our interest is
in the near-degenerate regime (rather than the Nariai
solution {\it per se}),  and the naive extrapolation
of non-degenerate thermodynamics does indeed
suggest a  limit of vanishing temperature.} 
\par
It may be bothersome that the first law of thermodynamics
is apparently violated at  the cosmological horizon,
$r_H =r_+$, given that the entropy is increasing
($\Delta S_{+}>0$)
while the mass is decreasing ($\Delta M<0$).
However, this  paradoxical feature of 
 cosmological horizons  can be anticipated and is, in fact, well understood 
\cite{SSV}. (Reassuringly, there is no such violation
for the black hole horizon, $r_H=r_-$.)
Speaking of the first law, one finds that the expected
relation of
\be
{\partial|\Delta S_H|\over \partial |\Delta M|}= {1\over \Delta T_{H}},
\label{8.55}
\ee
is not quite obtained; rather, it is off by a simple
numerical factor. This slight discrepancy can be viewed as
a consequence of treating each of the horizons as an isolated
system. Nonetheless, this  effect can be accommodated for 
with a simple rescaling: $\Delta M\rightarrow \Delta M (n-1)/(n+1)$.
Hence, we now have (for either horizon)
\be
|\Delta S_{H}| ={1\over\hbar} 
\sqrt{n+1\over n-1} 
\sqrt{{n \pi \V\over(n-1)}{r_0^{n+1}\over l^n }
|\Delta M|},
\label{7B}
\ee
\be
\Delta T_{H}= 2\hbar\sqrt{n-1\over n+1} 
\sqrt{{(n-1)\over n \pi \V}{l^n\over r_0^{n+1}}
|\Delta M|}.
\label{8B}
\ee

\section{The Effective  Theory}
\par
The next step in our procedure is  the dimensional reduction of the 
fundamental action (\ref{1})
to an effective two-dimensional theory \cite{REV}.
The reduction that we have in mind can be obtained by imposing
a spherical {\it ansatz}: 
\be
ds^2_{n+2}= ds_{2}^2(t,x)+ \phi^2(x,t) d\Omega_n^2,
\label{8.5}
\ee
which transforms Eq.(\ref{1}) into the following functional
 (also see \cite{newk,x5}):
\be
I= {\V\over 16 \pi l^n}
\int d^2x \sqrt{-g}\phi^{n}
\left[R+ n(n-1)\left({(\nabla \phi)^2\over \phi^2}
+{1\over \phi^2}\right)- {n(n+1)\over L^2}\right].
\label{9}
\ee
Here,  the ``dilaton'', $\phi$, can be identified with the radius of the 
symmetric $n$-sphere  and all 
geometric quantities are now  defined with respect to the resultant
 1+1-dimensional manifold.
\par
It proves to be convenient if the dilaton is  redefined
in accordance with
\be
{\psi}(x,t)= \left[{\phi\over l}\right]^{n\over 2}.
\label{9.5}
\ee
The reduced action (\ref{9}) can  then be reformulated as follows:
\be
I= {1\over 2G}
\int d^2x \sqrt{-g}
\left[D(\psi) R+ {1\over 2}\left(\nabla \psi\right)^2 
+{1\over l^2} V_{\Lambda}(\psi) \right],
\label{9.75}
\ee
where the following definitions have been invoked:
\be
{1\over 2G}\equiv {8(n-1)\V\over 16 \pi n},
\label{9.80}
\ee
\be
D(\psi)\equiv {n\over 8(n-1)}\psi^2,
\label{9.90}
\ee
\be
 V_{\Lambda}(\psi)\equiv
{n^2\over 8}\psi^{{2n-4\over n}}- {(n+1)n^2 l^2\over 8(n-1)L^2}\psi^2.
\label{9.95}
\ee
\par
The above  action is an particularly convenient form,
as it is now suitable for the implementation of a field reparametrization 
that  eliminates the kinetic term \cite{lk2}.
Following the prescribed methodology, let us redefine the dilaton, metric
and ``dilaton potential'' as follows:
\be
\Y=D(\psi)={n\over 8(n-1)}\psi^2,
\label{9A}
\ee
\be
{\overline g}_{\mu\nu}= \Omega^2(\psi) g_{\mu\nu},
\label{9B}
\ee
\be
{\overline V}_{\Lambda}(\Y)= {V_{\Lambda}(\psi)\over \Omega^2(\psi)},
\label{100}
\ee
where
\be
\Omega^2(\psi)\equiv \exp\left[{1\over 2}\int 
{d\psi\over \left(dD/d\psi\right)}\right] = {\cal C}\left[
{8(n-1)\Y\over n}\right]^{{n-1\over n}}.
\label{10}
\ee
In the last line,  ${\cal C}$ represents an arbitrary   constant of 
integration that  can (in a general sense)  often be 
fixed via physical arguments.
Here, we will follow  the fifth section  of  \cite{newk}
(which similarly considers the  spherical reduction
of an arbitrary-dimensional action) 
and  set  ${\cal C}= n^2/8(n-1)$.
\par
With  the above  reparametrizations, the reduced action (\ref{9.75})  
 transforms into  the following expression:
\be
I={1\over 2G}\int d^2x 
\sqrt{-{\overline g}}\left[{\overline \psi} R({\overline g})+ 
 {1\over l^2} {\overline V}_{\Lambda}({\overline \psi})\right ].
\label{11}
\ee
\par
It  can readily be seen  that the  
degenerate-horizon limit of the higher-dimensional theory
is   precisely recovered 
when  ${\overline V}_{\Lambda}({\overline\psi})=0$.
Denoting  this extremal value of the dilaton as   
${\overline \psi}={\overline \psi}_0$,
we find that ({\it cf}, Eqs.(\ref{100},\ref{9.95},\ref{9A}))
\be
{\overline \psi}_0
={n\over 8(n-1)}\left[(n-1)L^2\over  (n+1) l^2\right]^{n/2}.
\label{13}
\ee
\par
With the above discussion  in mind,
let us  define ${\tilde \psi}\equiv {\overline \psi}-{\overline \psi}_0$
and appropriately expand the action (\ref{11}) about the extremal
configuration. To first order in ${\tilde \psi}$, such an
expansion yields
\be
I={1\over 2G}\int d^2x 
\sqrt{-{\overline g}}\left[{\tilde \psi} R({\overline g})+ 
 {1\over l^2} {\tilde V}_{\Lambda}({\tilde \psi})\right ],
\label{14}
\ee
where
\bea
{\tilde V}_{\Lambda}({\tilde\psi})&\equiv&\left.
{d {\overline V}_{\Lambda} \over d{\overline\psi}}\right|_{{\overline\psi}_0}
{\tilde \psi} 
\nonumber \\
&=& 
-16{(n-1)^2\over n^2}\left[{(n+1)l^2\over (n-1) L^2}\right]^{(n+1)/2}
 {\tilde \psi}
\nonumber \\
&\equiv & -2\lambda  {\tilde \psi}.
\label{15}
\eea
Note that Eqs.(\ref{9.95},\ref{9A},\ref{100},\ref{10},\ref{13})
were  applied in attaining the second to last line
and  $\lambda>0$. 
\par
 We now drop the various tildes and bars and   thus consider
the following action:
\be
I={1\over 2G}\int d^2x \sqrt{- g}\psi\left[ R( g)- 
 2{\lambda\over l^2} \right ].
\label{16A}
\ee
We could, in principle, continue on with this form  of effective action;
essentially,  a two-dimensional de Sitter theory of
constant curvature  \cite{CAD}. 
However, it is substantially easier to work with the anti-de Sitter 
analogue,
for which the static solutions are described by the 
very well-known Jackiw-Teitelboim (JT) black hole \cite{jt}.
This  JT  analogue can be obtained  with a simple analytic continuation
of the length parameter ({\it i.e.},  $l^2\rightarrow -l^2$),  and
so we will subsequently work with 
\be
I={1\over 2G}\int d^2x \sqrt{- g}\psi\left[ R( g)+ 
 2{\lambda\over l^2} \right ],
\label{16}
\ee
where  $\lambda$ is (still) a positive quantity.
Actually, if the reduced action  described a fundamental theory,
then such a continuation would have to be treated
very carefully; especially,  in any attempt to  
extract physically meaningful results about the spacetime. However,
in the current analysis, where the reduced theory
serves as {\it only} an effective description of  a reduced phase space, 
 such a procedure should  be perfectly viable.
\par
The general  solution of this effective  JT  action (\ref{16})
can be conveniently expressed in a static, Schwarzschild-like
gauge:
\be
ds^2= -(\lambda {x^2\over l^2}-2lGm)dt^2+(\lambda {x^2\over l^2}-2lGm)^{-1}
dx^2,
\label{17}
\ee
\be
\psi={x\over l}.
\label{18}
\ee
Here,  $m$ represents the  conserved mass of the JT black hole.
Moreover, we can apply the formalism of \cite{lk3}
(which is applicable to a  generic theory of two-dimensional dilaton  gravity),
to obtain the other  thermodynamic properties
of relevance; namely, the entropy and the temperature:
\be
S_{JT}={2\pi\over \hbar G} \psi_{+},
\label{19}
\ee
\be
T_{JT}={\hbar\lambda\over 2\pi l}\psi_{+},
\label{20}
\ee
where $\psi_{+} = x_{+} /  l =\sqrt{2lG m / \lambda}$
is the horizon value of the dilaton field.
\par
Substituting for $\lambda$ (\ref{15}) and 
$G$ (\ref{9.80})
  into the above expressions, as well as 
applying the defining relation for $r_0$ (\ref{2.90}),
 we are able to derive the following results:
\be
S_{JT} ={1\over\hbar}  
\sqrt{{n \pi \V\over(n-1)}{r_0^{n+1}\over l^n } m},
\label{21}
\ee
\be
T_{JT}= 2\hbar 
\sqrt{{(n-1)\over n \pi \V}{l^n\over r_0^{n+1}}
m}.
\label{22}
\ee
\par
Let us now 
 make a  reasonable identification between 
the JT black hole mass, $m$, and the mass deviation (from
degeneracy)
of the higher-dimensional model,  $|\Delta M|$. It then becomes
quite evident that 
the two theories are closely related. To be explicit
({\it cf}, Eqs.(\ref{7B},\ref{8B})),
\be
S_{JT}={\cal K}|\Delta S_{H}|,
\label{23}
\ee
\be
T_{JT}={1\over {\cal K}}\Delta T_{H},
\label{24}
\ee
where ${\cal K}=\sqrt{n-1/n+1}$ is a dimensionless numerical factor.
Clearly, we can eliminate
the numerical factor, ${\cal K}$, 
by  rescaling  an appropriate
parameter of the  effective theory
(such as the {\it effective} Planck's constant).
The important point is that $T_{JT}S_{JT}=\Delta T_{H}|\Delta S_{H}|$ 
and all of the  dimensional
quantities do indeed coincide precisely.
Hence, we have established the anticipated thermodynamic duality:
the  nearly degenerate sector of Schwarzschild-de Sitter black holes
 with the  near-massless sector of JT theory.
\par
Before proceeding to the next section,
let us point out an important  property of
 the JT black hole: as the mass, $m$, goes to zero,
so does the temperature, $T_{JT}$, and the entropy, $S_{JT}$. 
Although this relation
is intuitively expected in {\it conventional} thermodynamic systems,
it is, nevertheless,  
very unusual in a black hole context (where, typically, a
vanishing temperature corresponds to both a finite
mass and entropy).
It is, in fact, this
property of JT black holes that makes them an ideal
framework for studying the low-temperature regime of
dually related theories.

\section{Quantum Effects} 
\par
In analogy to our prior studies on near-extremal black
holes  \cite{new,MED}, we will now invoke one-loop considerations to
 argue the following:
 a nearly degenerate Schwarzschild-de Sitter solution will be unable
to  evolve continuously into a degenerate,  Nariai
spacetime (and {\it vice versa}). 
\par
The essence of our  argument goes as follows (for a more
detailed discussion, see \cite{new}).
We start by considering the simplest possible matter having a 
higher-dimensional
pedigree; namely,
a  massless scalar field ($f$) that
is minimally coupled to the $n$+2-dimensional de Sitter theory.
 The revised
(total) action   can now be expressed as
\be
I^{(n+2)}_{TOT}=I^{(n+2)}-{\hbar\over 16\pi l^n}\int d^{n+2}x\sqrt{-g^{(n+2)}}
(\nabla^{(n+2)}f)^2,
\label{32}
\ee
where $I^{(n+2)}$ is the classically defined  action of Eq.(\ref{1}). 
Again imposing the  reduction {\it ansatz} of 
Eq.(\ref{8.5}), as well as
constraining  $f=f(t,x)$, we obtain the following   reduced 
form:
\be
I_{TOT}=I-{\hbar\V\over 16\pi l^n}\int d^2 x\sqrt{-g}
\phi^n (\nabla f)^2,
\label{33}
\ee
where $I$ is the reduced action of Eq.(\ref{9}).
\par
Following the  same pattern of field reparametrization and expansion as
described in Section 3, we can eventually write  
\be
I_{TOT}=I_{JT}-{\hbar(n-1)\V\over 2\pi n}\int d^2 x\sqrt{-{\overline g}}
{\tilde \psi}
({\overline\nabla}f)^2,
\label{WEEE}
\ee
where $I_{JT}$  is the JT action of  Eq.(\ref{16}),
and  the tilde and bar notation  has been resurrected
for maximal clarity.
\par
The above result informs us that
the dilaton-matter coupling  is precisely
that obtained in the dimensional reduction (from three to two dimensions) 
of a BTZ black hole;  under the assumption of
 minimally coupled matter in the three-dimensional
theory \cite{btz,ao}.\footnote{The unorthodox  factor
in front of the above matter action is of no relevance, as this constant
can always be
absorbed in a rescaling of $f$.}
This observation is important because,
as  discussed on a  rigorous level in \cite{new}, the dimensionally
reduced BTZ black hole suffers a formal breakdown as
the limit of zero mass is approached ({\it i.e.}, as $m\rightarrow 0$).
More specifically, at some small but  finite value of $m$, the surface
gravity (or, equivalently, the temperature) will take
on a negative value and  become increasingly more
negative as the $m=0$ limit is approached.
Significantly, it has been argued that a non-negative surface gravity
is a necessary criteria for a consistent black
hole solution \cite{aht}.
On this basis, we have further argued that, 
because of quantum back-reaction effects, the
reduced BTZ black hole   will    
be unable to attain a state of vanishing mass \cite{new}.
Obviously, the same conclusion (if valid) must apply to
the JT black hole when it is coupled to matter with a higher-dimensional
pedigree.
\par
Let us now remind ourselves of the main outcome of Section 3;
namely, the near-massless limit of the JT sector is dual
(at least at the level of thermodynamics) to
the near-degenerate sector of a  Schwarzschild-de Sitter black
hole. It therefore stands to reason that  the censorship
of   massless  JT black holes
will translate into the following dual statement: a Schwarzschild-de Sitter
black hole will be unable to reach the limit of horizon degeneracy
({\it i.e.}, the limit  $|\Delta M|\rightarrow 0$). 
To put it another way, quantum back-reaction effects
will prohibit a nearly degenerate Schwarzschild-de Sitter
black hole from ever reaching a precise state of
degeneracy. Hence, a non-degenerate Schwarzschild-de Sitter
spacetime should never be able to evolve continuously into
a Nariai black hole and {\it vice versa}.

\section{Conclusion} 
\par
In summary, we have been considering  a  holographic type of duality
\cite{x5}
between a nearly degenerate 
Schwarzschild-de Sitter spacetime (of arbitrary dimensionality) and
the near-massless sector of Jackiw-Teitelboim theory \cite{jt}.
We started out by rigorously demonstrating this dual
relationship at the level of horizon thermodynamics.
We then went on to argue that, on the basis of
this duality and one-loop considerations, the higher-dimensional
spacetime will never be able to continuously evolve into
a Nariai solution \cite{NAR}; that is, a Schwarzschild-de Sitter
black hole of maximal mass and perfectly coincident horizons.
Let us emphasize that our conclusions in no way undermine
the  Nariai black hole as a viable
solution of a spacetime with a positive cosmological constant. Moreover, 
the type of censorship discussed here does not apply to
more exotic mechanisms, such as quantum tunneling 
between Nariai and non-degenerate solutions \cite{BOU}.  
\par
It should again be pointed out that the latter part of the paper was
based (vicariously through \cite{new,MED}) on  the outcomes of a  one-loop
 analysis of the  dimensionally reduced BTZ black hole.
This analysis was based, in turn, on  a form
of  dimensionally-reduced action  
that has been the subject
of some recent criticism \cite{FRO,ZEL}. In spite of
the potentially  disturbing implications, 
we expect that our qualitative outcomes  
will not be in any jeopardy. To motivate
this claim, let us take note of some independent
studies in the context of black hole spectroscopy \cite{das,US}.
In these treatments, it was shown, quite generically,  that 
quantum fluctuations
will prevent   a charged or spinning
black hole  from ever reaching a state of exact extremality.
Although these studies are not directly applicable to the current analysis, 
their generality does insinuate
a similar censorship in any spacetime with
potentially degenerate horizons. 
\par
Finally, one might wonder if this discontinuity
between non-degenerate and degenerate Schwarzschild-de Sitter
black holes could have any ramifications in quantum gravity.
In this regard, let us take note of the
conjectured duality  between time evolution in an 
asymptotically de Sitter spacetime and a suitably defined 
renormalization-group flow \cite{STR2}. (This duality can be viewed
as a manifestation of the so-called ``dS/CFT'' holographic 
correspondence \cite{STR1}.) Perhaps significantly, 
it has been suggested that
the Nariai solution should correspond to  the infrared
fixed point of the holographic flow \cite{HAL}. It is not obvious
to us, at the current time, if our findings could play a 
role in such a  context; nonetheless, it could yet prove to be a
worthwhile direction of investigation.

\section{Acknowledgments}
\par
The author would like to thank V.P. Frolov for helpful conversations.
\par

\end{document}